\documentclass[Afour,times,sagev]{sagej}

\usepackage{moreverb,url}
\usepackage{natbib}
\usepackage[colorlinks,bookmarksopen,bookmarksnumbered,citecolor=red,urlcolor=red]{hyperref}
\pdfoutput=1
\newcommand\BibTeX{{\rmfamily B\kern-.05em \textsc{i\kern-.025em b}\kern-.08em
T\kern-.1667em\lower.7ex\hbox{E}\kern-.125emX}}

\def\volumeyear{2016}

\begin{document}

\runninghead{Kova{\v c}evi{\'c}}

\title{Two-Dimensional (2D) Hybrid Method: Expanding 2D Correlation Spectroscopy (2D-COS) for Time Series Analysis}

\author{Andjelka B. Kova{\v c}evi{\'c}}

\affiliation{University of Belgrade-Faculty of Mathematics, Department of Astronomy,
Studentski trg 16, 11000 Belgrade, Serbia}

\email{andjelka.kovacevic@matf.bg.ac.rs}

\begin{abstract}
We present a concise report on the '2DHybrid' method, an innovative extension of two-dimensional correlation spectroscopy (2D COS), tailored for quasar light curve analysis. Addressing the challenge of discerning periodic variations against the background of intrinsic "red" noise fluctuations, this method employs cross-correlation of wavelet transform matrices to reveal distinct correlation patterns between underlaying oscillations, offering new insights into quasar dynamics.

\end{abstract}

\keywords{2D correlation spectroscopy, red noise, binary quasars}

\maketitle

\section{Introduction}

{
Black holes, phenomena of extreme density and gravity from which even light cannot escape, are not only central to our understanding of the universe but also present unique challenges and opportunities for spectroscopic analysis. These enigmatic objects vary in size, from stellar black holes formed from collapsing massive stars to supermassive black holes, millions of times heavier than our Sun, residing in the centers of almost every galaxy, including the Milky Way. A key phenomenon associated with supermassive black holes is the quasar ('quasi-stellar object'), which results from the interaction and accretion of matter around these black holes. Quasars, once mistaken for stars due to their intense brightness, are crucial for studying galactic dynamics and evolution.
Spectroscopy, an integral tool in astrophysics, is invaluable for investigating these phenomena, particularly in analyzing the behavior and composition of matter around quasars. Through spectroscopic analysis, we can gain insights into the light emissions of quasars, which exhibit variability across the electromagnetic spectrum.  A key area of interest in current research is the fate of black holes and the resultant quasars when galaxies collide and merge, forming binary systems (see Figure \ref{fig:figure1}) whose complex dynamics are still not fully understood. Quasars exhibit variability across electromagnetic spectrum, a phenomenon where their brightness fluctuates over time. This variability ($X(t)$) is stochastic,   such that their power
spectrum density $\Gamma _{X}(\nu)\sim |\nu|^{-\alpha}$. It is often statistically modeled using a 'damped random walk', a concept defined by an exponential covariance
matrix that behaves as a random walk for short time scales and asymptotically achieves a finite
variability amplitude at long time scales.  The advent of large-scale astronomical surveys has led to an unprecedented accumulation of quasar light curves — records of integrated  flux measurements over time. These extensive datasets open new avenues for a systematic and detailed exploration of quasar variability, particullarly binary quasar candidates. 
}

In the advancing field of space-time physics, the significance of electromagnetic indicators, particularly in zeroth order optical moments, is increasingly recognized for identifying nearby binary quasar candidates. \citep[][]{THEZA} Periodic variability, as indicated by magnetohydrodynamic simulations, has emerged as a key marker \citep{Guti_2022} for the presence of these elusive objects. Recent large time-domain surveys have identified $\sim 250$ binary quasar candidates \citep{dorazio23}, many of which are targeted for observation with next-generation interferometers. \citep{ngeht23} The quasar  emission resembles red noise, displaying fractal-like, self-similar patterns \footnote{Self-similarity (or self-affinity) means that the reperesentation of stochastic signal $(t, X(t))$ remains statistically unchanged when the time axis and the amplitude are simultaneously scaled by some factor.} across various time scales. \citep[][]{vio91,belete18} Searching for any coherent signal buried within the red noise and in combination with unfavorable sampling, confronts a significant challenge: the Fourier uncertainty principle. Analogous to the Heisenberg principle in quantum mechanics, this principle in signal processing is known as the Gabor limit \citep{gabor47}, delineating the constraints on precisely measuring a signal's instantaneous frequency. Time-frequency analysis has demonstrated that linear operators cannot breach this uncertainty bound, necessitating a nonlinear approach to overcome the Gabor limit. \citep{cohen1995time} In response, we have extended 2D Correlation spectroscopy \citep[2DCOS,see][]{NODA20,noda19, noda93gen, noda20pro,noda24adv,noda07two,noda08rec,noda10pro,noda14fro,noda14froB,noda14two, noda14vib,noda93rec} idea to the integrated spectrum of quasars over long time baselines, which is an innovative nonlinear methodology specifically designed to surpass the Gabor limit.
{
Our method involves principles of 2D COS: cross-correlating the wavelet transform of quasar integrated emissions, commonly referred to as light curves, to obtain a correlation plane of detected oscillations. These wavelet-based estimators, ideally suited for the spectrum analysis of stochastic processes, are effective in handling non-stationary data. \citep{Abry1995} Moreover, in spectral analysis, wavelets have demonstrated significant potential in various chemical studies for noise removal, resolution enhancement, data compression, and chemometric modeling. \citep[][]{chemical} 
Therefore, our approach demonstrates the conjugation of wavelet transform and 2D COS principles, suggesting a broader applicability in the field of spectroscopy.}
Here we report on the current progress of our method for nonlinear processing and analyzing quasar light curves, tailored for the  large time domain astronomical survey Vera C. Rubin Legacy Survey of Space and Time. \citep[][]{ivezic2019}

\section{Challenges in quasar periodicity mining}

\begin{figure}
\centering
\includegraphics[width=.45\textwidth]{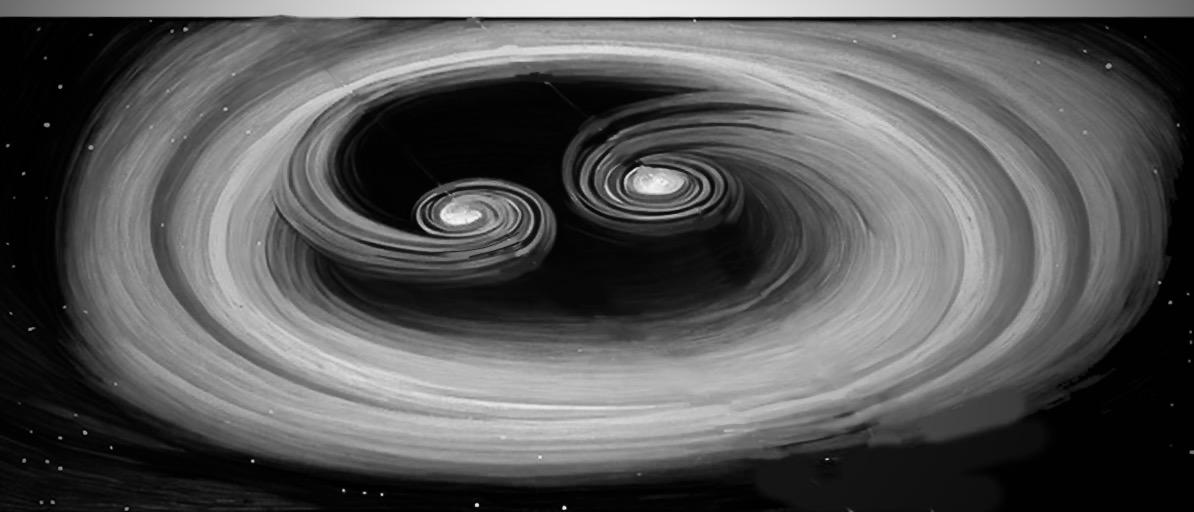}
\caption{Artistic rendition of a SMBBH system showcasing a triple disk structure at subparsec scales. This visualization is based on simulations by NASA's Goddard Space Flight Center/Scott Noble, available at \url{https://svs.gsfc.nasa.gov/vis/a010000/a013000/a013043/13043_SMBH_Simulation_1080.mp4}.Central to the image are the two black holes surrounded by a circumbinary disk that illustrates the dynamic gas transfer within the binary system and the formation of individual accretion disks encircling each black hole.}
\label{fig:figure1}
\end{figure}

{Understanding the periodicity of quasars is crucial in astrophysics, particularly in the context of galaxy evolution and binary supermassive black hole (SMBH) systems. The initial stages of galaxy collisions at kiloparsec scales, characterized by dual quasars each hosting an active SMBH, have been observed across various wavelengths. \citep{DEROSA2019} However, detecting subparsec (with separations below 3.26 light years) binary quasars presents significant challenges due to the limitations in resolving such close separations. \citep{DEROSA2019} These cosmic events often trigger substantial gas inflows to the central regions of galaxies, suggesting that binary quasars are likely enveloped in dense gas. This accretion process can lead to electromagnetic emissions from the SMBHs, allowing their indirect detection through interactions with the surrounding gas. \citep{bogdanovic22}
To detect sub-parsec binary quasars, astronomers primarily rely on two methods. Spectroscopy becomes useful when significant gas surrounds one or both SMBHs, as shown in Figure \ref{fig:figure1}. This surrounding gas produces broad emission lines, with the orbital motion of the SMBHs manifesting as Doppler shifts in these lines. \citep{dorazio23} At smaller separations, like milli-parsecs where broad emission lines are truncated, periodic fluctuations in the photometric light curves of quasars become key indicators. These fluctuations might originate from gas streams periodically accreting from a circumbinary disk onto the SMBHs, as illustrated in Figure \ref{fig:figure1}, or from Doppler-boosted emissions due to the relativistic motion of gas around each SMBH. In such binary systems, the SMBHs can orbit at velocities up to a few percent of the speed of light.
Periodic perturbations in the circumbinary disk can modulate the accretion rate, leading to observable variations in the light curve that typically align with the binary's orbital period. This is especially true in binaries where the mass ratio of the smaller to larger mass component lies within the range $[0.05, <0.3-0.5]$. Moreover, significant Doppler boost variability is expected, resulting in a coexistence of periodic accretion and relativistic Doppler boost effects. If Doppler boosting is dominant, the variability in the light curve would likely be smooth and quasi-sinusoidal.
Interestingly, none of the identified binary candidates are close to their final coalescence, meaning that the observed periods in light curves remain constant over time. However, a recent study introduced the first system, J143016$+$230344 \citep[see][]{Masterson_2023}, exhibiting a rapidly decaying period, as evidenced in its optical and X-ray light curves. The period has shortened from approximately one year to one month within just three years. This diminishing period signal, termed the 'tick tock' signal $S(t)$, represents a damped sinusoid, characterized by gradual reductions in both amplitude and frequency:
\begin{equation}
S(t)=e^{-d_a t} \sin\left(2 \pi \frac{t}{P_{0} e^{d_f t}}\right)
\label{eq:damped}
\end{equation}
\noindent where:
\begin{itemize}
\item $t$ is the time variable,
\item $d_a$ is the amplitude damping factor,
\item $d_f$ is the frequency damping factor, and
\item $P_0$ is the initial period of the undamped signal.
\end{itemize}
Therefore, the 'tick tock' signal serves as a  proxy for understanding the complex dynamics associated with merging SMBHs within binary systems. Detecting periodic or damped periodic signals in quasar light curves, essential for identifying subparsec binary quasars, is accompanied with challenges. \citep[][]{dorazio23, DEROSA2019} The primary difficulty arises from the inherent nature of quasar emissions, which typically exhibit 'red noise' or power law spectra ($1/f$ processes, where $f$ is frequency). This type of signal complicates the identification of true periodicity due to its stochastic and variable nature.
Another significant challenge is the inherent limitations of current observational techniques. The periods of interest often exceed the timescales over which data have been collected, meaning that we cannot observe enough cycles to confirm periodicity definitively. Additionally, the data obtained from optical time-domain surveys are relatively sparse, further complicating the detection of periodic patterns.
Given that quasars' light curves typically exhibit red noise characteristics or power law spectra, wavelet-based estimators have been shown to correspond well to the analysis of such processes.\citep{Abry1995} Wavelets can effectively handle the variability and nonstationary nature of quasar emissions, offering a promising approach for analyzing them. }

\section{2DHybrid Method: A Novel Approach in Quasar Periodicity Analysis}
The periodogram serves as a tool to transform a one-dimensional signal from the time domain into the frequency domain. This transformation is particularly effective when dealing with stationary frequency spectra, where the frequencies present in the signal remain constant over time. However, in the context of quasar observations, this approach encounters limitations, as the signals we often observe are inherently non-stationary. Wavelet transforms, in contrast, offer a more nuanced conversion, turning a one-dimensional signal into a two-dimensional representation encompassing both time and frequency (or scale). Through scaleogram, we can glean  details about the system's state-space, shedding light on its dynamic behaviors.
Our  '2DHybrid' method represents a novel approach in this area. Building upon the foundational work in two-dimensional correlation spectroscopy (2D COS) developed by Isao Noda \citep{NODA20,noda19, noda93gen, noda20pro,noda24adv,noda07two,noda08rec,noda10pro,noda14fro,noda14froB,noda14two, noda14vib,noda93rec}, the 2DHybrid method \citep{2018MNRAS.475.2051K, 2019ApJ...871...32K, 2020MNRAS.494.4069K, 2020OAst...29...51K, 2023AJ....165..138F} effectively addresses the challenges posed by the non-stationarity of quasar light curves. Furthermore, the versatility of this method extends beyond astrophysics, finding applications in detecting oscillatory patterns in remote sensing of the upper atmosphere through very low frequency waves. \citep{mathKovacevic}

\subsection{Description of 2DHybrid method}

Distinct from traditional Fourier methods, the 2DHybrid method transforms signals from the one-dimensional time domain into a two-dimensional domain of period correlations, using the cross-correlation of wavelet matrices. While our method is compatible with a variety of wavelet transforms, in this analysis, we specifically employ the Weighted Wavelet Z-transform (WWZ). \citep{1996AJ....112.1709F} WWZ has been proven to be particularly effective in uncovering underlying periodicities in quasar light curves, making it an ideal choice for our purposes.

The cross-correlation of wavelet transforms of two signals 
$x(t)$ and  $y(t)$ is given by:
\begin{equation}
C(a, b) = \int_{-\infty}^{\infty} W_x(a, b) \cdot W_y^*(a, b) \, db
\end{equation}

\noindent where \( W_x(a, b) \) and \( W_y(a, b) \) are the wavelet transforms of \( x(t) \) and \( y(t) \) respectively, at scale \( a \) and translation \( b \). Here, \( W_y^*(a, b) \) denotes the complex conjugate of \( W_y(a, b) \).
 The nonlinearity of this operation stems from the product of the coefficients and the integration over the translation parameter $b$.

Given the WWZ transform frequency ($f$) and time ($\tau$) data, we can calculate when the Gabor limit is surpassed as 
\begin{equation}
\hat{GL} = \Delta \tau \cdot \Delta f < \frac{1}{4\pi}
\label{eq:glimit}
\end{equation}
\noindent with $\Delta f_{i} = f_{i+1} - f_{i}$ and $\Delta f = \frac{1}{m}\sum_{i=1}^{m-1} \Delta f_{i}$, as well as $\Delta \tau_{i} = \tau_{i+1} - \tau_{i}$ and $\Delta \tau = \frac{1}{n}\sum_{i=1}^{n-1} \Delta \tau_{i}$, where $m$ and $n$ are the number of frequency and time points, respectively.
{In this article, we demonstrate the robustness and versatility of the 2DHybrid method in detecting and characterizing periodic signals within quasar light curves, a capability we have previously established through various studies. \citep{2018MNRAS.475.2051K, 2019ApJ...871...32K, 2020MNRAS.494.4069K, 2020OAst...29...51K, 2023AJ....165..138F} Specifically, we present two distinct test cases to showcase the method's effectiveness. The first test case involves injecting a controlled artificial sinusoidal pattern into a synthetic quasar light curve, simulating simple periodic variability. The second test case applies the method to an observed u-band light curve data, into which we introduce a 'tick tock' signal - a representation of a damped sinusoidal pattern. These test cases effectively illustrate how the 2DHybrid method can discern and analyze both straightforward periodic and more complex damped sinusoidal variabilities in quasar emissions.}

\section{Results: test cases}
In this section, we present two test cases that demonstrate the application and effectiveness of the 2DHybrid method. These cases include controlled simulations, where we have injected a known sinusoid signal into synthetic quasar light curve, as well as real-world scenarios involving u-band light curve data from a quasar combined with damped sinusoids. Each test case serves not only to illustrate the method's capability in detecting and analyzing periodic signals but also to showcase its adaptability and robustness across different conditions and datasets
\subsection{Test Case 1: Synthetic quasar light curve and simple sinusoid Signal}
\begin{figure}
\centering
\includegraphics[width=.45\textwidth]{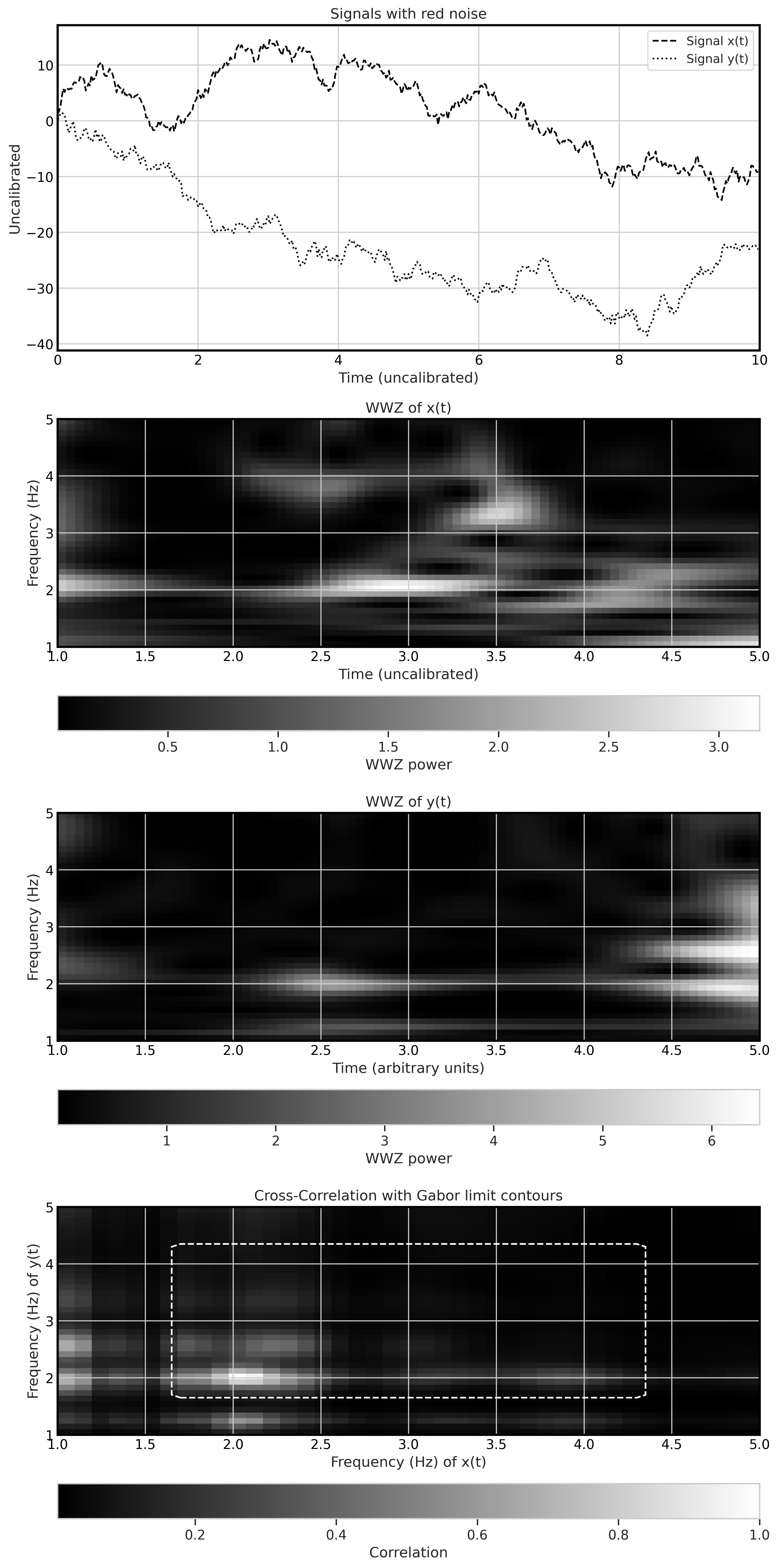}
\caption{Illustration of 2D Hybrid method on simulated quasar light curves. From top to bottom: the simulated light curves of quasars where x(t) and y(t) contain a signal of 2Hz; the WWZ transform of light curve x(t); the WWZ transform of light curve y(t); and a 2D hybrid correlation map of the detected oscillations in both light curves, overlaid with contours delineating the Gabor limit. The colorbars for the WWZ transforms represent the WWZ power, while the colorbar for the 2D Hybrid method reflects the correlation between the detected oscillations. }
\label{fig:figure2}
\end{figure}

The first test case of application of 2D Hybird method is given in Figure \ref{fig:figure2}. The top plot presents two synthetic time series datasets, labeled 'Signal x(t)' and 'Signal y(t).' These signals are composed of red noise, accounting for 70\% of the signal, and a sinusoidal component constituting the remaining 30\%. 'Signal x(t)' is depicted by a dashed line, and 'Signal y(t)' by a dotted line, with both incorporating a periodic signal at 2Hz. The second and third plots display the Weighted Wavelet Z-transform (WWZ) of 'Signal x(t)' and 'Signal y(t),' respectively. The grayscale intensity within these plots indicates the WWZ power, with brighter areas signifying greater power. The x-axis denotes time in arbitrary units, while the y-axis denotes frequency in Hz.

The expected 2Hz signal is  represented as bright spots or stripes, standing out against the background, owing to a sufficient signal-to-noise ratio. Red noise, however, tends to produce a smearing effect across the frequency domain in the WWZ power plot, potentially rendering peaks that signify the actual signal less distinct.

The bottom plot illustrates the cross-correlation of the WWZ transforms of 'Signal x(t)' and 'Signal y(t),' as shown in the preceding subplots, expanded on a frequency-frequency plane. Here, the grayscale  reflects the cross-correlation magnitude, with brighter areas indicating stronger correlations. Since red noise generally exhibits a lack of correlation between distinct signals, its influence is diminished in the cross-correlation plot, which facilitates a more unambiguous identification of shared periodic components.

Contoured region on the cross-correlation map is a zone where the 2D Hybrid method attains an enhanced resolution, surpassing the conventional limits defined by the Gabor limit $1/ (4\pi )$ (see Eq. \ref{eq:glimit}). These contours, calculated via a morphological closing operation, serve to not only delineate areas with significant periodic signal correlation but also to augment interpretability by amalgamating proximal patches into unified regions. This visualization strategy enables the distinction of true periodicities emmersed in the red noise background, a particularly advantageous feature for the analysis of astronomical time series. The detection of pronounced correlation peaks at the anticipated frequencies, notwithstanding the pervasive red noise, validates the efficacy of the 2D Hybrid method, as an extension fo 2D COS, in isolating signals from noisy astronomical data.

Furthermore, the 2D Hybrid method, through 2D COS method inherits the wavelet transform's capability for scaling and shifting based on the signal's frequency content. Combined with the nonlinear nature of the cross-correlation process, this approach is inherently nonlinear. Such a nonlinear methodology has the potential to furnish a more granular time-frequency analysis of the signals, thereby potentially overcoming the restrictions imposed by the Gabor uncertainty principle on linear analyses.

\subsection{Test case 2: Artificial tick tock signal and real observed quasar light curve }

\begin{figure}
\centering
\includegraphics[width=.45\textwidth]{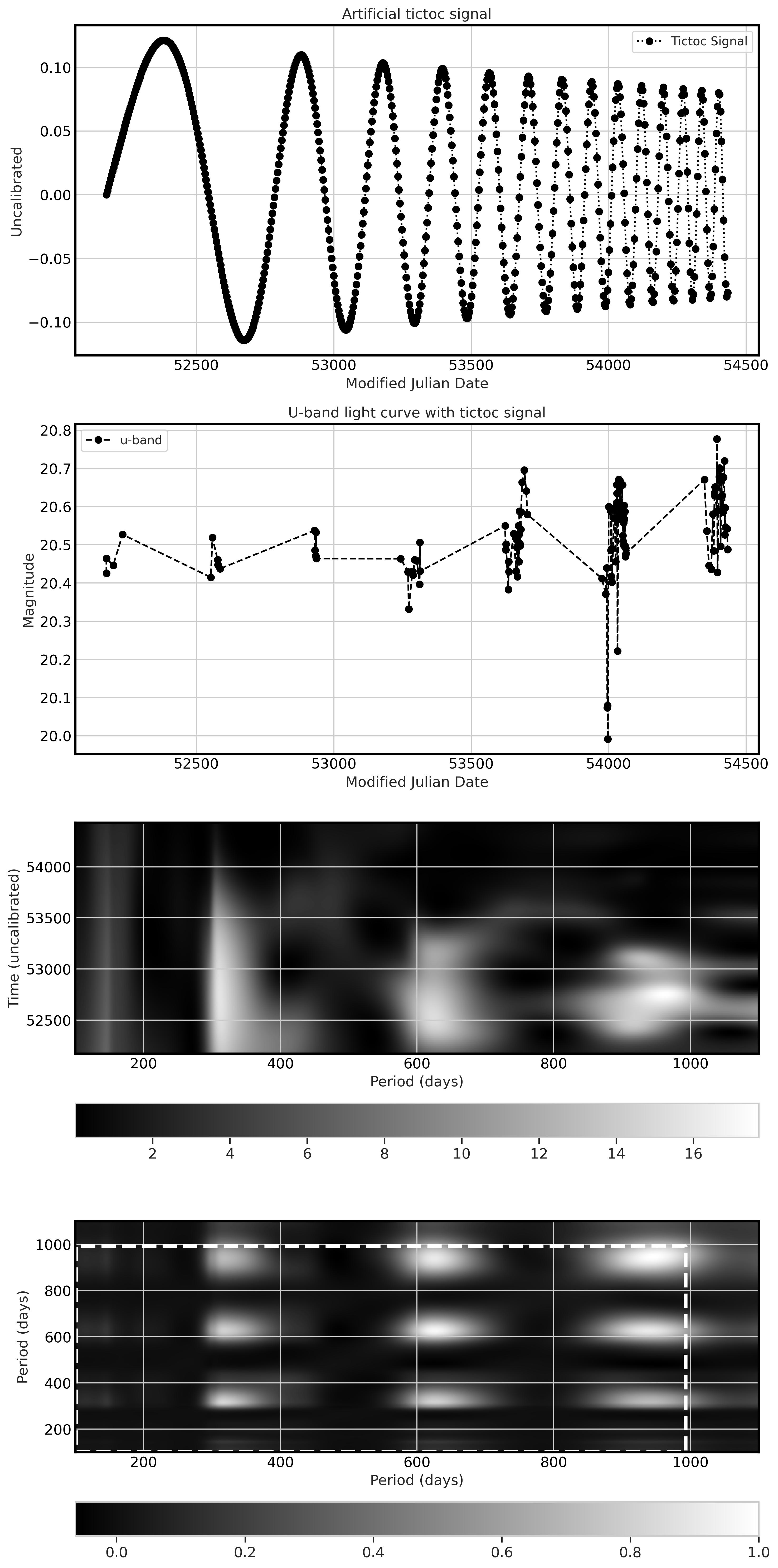}
\caption{Application of 2D Hybrid method on an observed quasar light curve with artificially injected tick tock signal. From top to bottom: the artificial tick tock signal with an initial period of 1000 days; the real u-band light curve of a quasar with the artificial damped sinusoid signal injected, dots are observed points, dashed line is eye-guiding; WWZ power map of the u-band light curve with artificially incjected tick tock signal;  and a 2D hybrid autcorrelation map of the detected oscillations in the quasar light curve, overlaid with contours delineating the Gabor limit. The colorbars for the WWZ transforms represent the WWZ power, while the colorbar for the 2D Hybrid method reflects the correlation between the detected oscillations. }
\label{fig:figure3}
\end{figure}

The top panel of  Figure \ref{fig:figure3} illustrates the artificial tick tock signal (see Eq.\ref{eq:damped}), a sine wave with  an initial period of $P_{0}=1000$ days, subject to damping in both amplitude and frequency with factors $d_{a}=0.0002, d_{f}=0.0008 $, respectively. This damping emulates the dissipative processes of merging two supermassive black holes that may occur in actual astrophysical settings. The signal was inserted into the observed u band light curve of quasar 1460382 from the LSST AGN Data Challenge database \citep{yu_22}, with a moderate signal to noise ratio $\sim 1.5$.

The second panel  shows the actual u-band light curve from a quasar, now with the tick tock signal embedded within it. 
The WWZ power spectrum of the modified light curve (third panel) exhibits the anticipated power enhancement corresponding to the tick tock signal's period. Despite the red noise, which tends to smear the spectral features, the WWZ analysis successfully delineated the periodic signal, highlighting its power at different timescales.

The fourth panel presents the mapping of the autocorrelated WWZ power onto a period-period plane, which allowed for a visualization of the cross-correlation of the light curves' periodic components. Crucially, we overlaid contours on this plane where the product of the time and frequency resolutions was smaller than the Gabor limit.  This delineation is crucial as surpassing the Gabor limit implies that the observed periodicities within these regions are resolved with a precision beyond the conventional bounds set by linear time-frequency analysis methods. Consequently, these contours underscore the presence of potentially meaningful astrophysical signals.  Notably, the brightest areas on the map correspond to the known parameters of the injected tick tock signal, validating our approach. The presence of these bright regions amidst the darker background confirms that the periodic signal has been effectively isolated from the noise.

The WWZ power spectrum for a simple sinusoid shows a distinct bright stripe or area corresponding to the 2Hz period (see second and third plots in Figure \ref{fig:figure2}, whereas the tick tock signal  creates a more spread out or smeared power spectrum (third subplot in Figure \ref{fig:figure3}) due to its changing characteristics.
In the case of the simple sinusoid, the cross-correlation map would likely show a strong correlation peak at 2Hz, with less complexity in the contours (fourth subpanel in Figure \ref{fig:figure2}. For the tick tock signal, the cross-correlation map  displays off diagonal patterns, reflecting the signal's complex damping nature (fourth subpanel in Figure \ref{fig:figure3}).

\section{Discussion and Summary}

The two test cases presented  effectively demonstrate the versatility of the 2DHybrid method in analyzing quasar light curves. Through these cases, we have shown how the method adeptly handles both simple sinusoidal signals and more complex, dynamically changing signals like the tick tock signal.

The WWZ power spectrum of a simple sinusoid, as seen in the second and third plots of Figure \ref{fig:figure2}, presents a distinct bright stripe or area corresponding to the 2Hz period. Here the smearing effect is attributed to the presence of the red noise.  On the other hand, the tick tock signal, which simulates the characteristics of merging supermassive black holes, exhibits a more spread out or smeared power spectrum (as shown in the third subplot of Figure \ref{fig:figure3}). These smearing effects are attributed to the red noise and signal's damped nature in both amplitude and frequency, presenting a more challenging detection scenario.

Interestingly, the cross-correlation maps of WWZ offer more precise localization of periodic signals. For the simple sinusoid, we observe a strong correlation peak at 2Hz with less complex contours (fourth subpanel of Figure \ref{fig:figure2}), whereas the tick tock signal generates off-diagonal patterns on the map, reflecting its complex damping nature (fourth subpanel of Figure \ref{fig:figure3}). These patterns are crucial in distinguishing the simple sinusoid signal of pre merging scenarios  from those close to the merger, showcasing the 2DHybrid method's nuanced detection capabilities.

Overall, our analysis underscores the 2DHybrid method's efficacy in extracting meaningful periodic signals from noisy datasets, a task of paramount importance in astrophysical research. The method not only excels in identifying clear-cut periodicities but also demonstrates its prowess in unraveling the complexities of dynamically evolving signals. This blend of precision and adaptability positions the 2DHybrid method as a valuable tool in the ongoing time domain  study of quasars.

{Our approach signifies an innovative integration of wavelet transform and 2D COS, indicating its substantial potential for wider applications in spectroscopy, especially as  wavelets have been increasingly used in spectral analysis, especially in chemical studies.}

\begin{acks}
The author extends sincere thanks to the Reviewers for their invaluable comments. Additionally, the author is grateful to the Organizers of the 12th International Symposium on Two-Dimensional Correlation Spectroscopy, held on August 17-18, 2023.  A.B.K. acknowledges funding provided by the University of Belgrade - Faculty of Mathematics (the contract 451-03-66/2024-03/200104) through the grants by the Ministry of Science, Technological Development and Innovation of the Republic of Serbia. 
\end{acks}

\bibliographystyle{SageV}
\bibliography{sage}
\end{document}